# Modified Computation of Correlation Integral for Analyzing Epileptic Signals

Prajna Upadhyaya, Tohru Yagi

*Department of Mechanical Engineering, Tokyo Institute of Technology, 2-12-1 Meguro, Ookayama, Tokyo, 〒152-8550*

***Abstract:*** Epilepsy is a chronic neurological disorder characterized by recurrent seizures. One method for analyzing seizure activity is to compute the correlation dimension of time-series electroencephalographic signals. The Grasserberg and Proccacia algorithm is commonly used to compute this correlation dimension. The algorithm uses the Heaviside function to determine the correlation integral by counting the number of distances between vectors ($d_{ij}$) that are greater than a threshold. However, information about the chaotic nature of the signal is not completely retained by this function. In this work, instead of using the Heaviside function, we calculated the correlation integral by using an exponential function of $d_{ij}$. Greater sensitivity to the interictal and ictal signals using this modified algorithm was verified using three datasets. Comparing heatmaps of $d_{ij}$ obtained using the original and modified methods showed additional information that was retained with the new algorithm.

## 1. INTRODUCTION

Neurological disorders affect millions of people worldwide. Epilepsy is one such disorder and affects almost 50 million people globally [1]. Although epilepsy is not a critical condition compared with other neurological disorders, it can be life threatening if not diagnosed promptly and treated appropriately. Uncontrolled seizures are one of the most dangerous symptoms of epilepsy and can be life threating if they occur when the person is alone or performing daily activities such as driving, crossing the street, or cooking. Although antiepileptic drugs are available, they are ineffective for more than 30% of people with epilepsy [2]. In addition, prolonged use of antiepileptic drugs can lead to cognitive and neurological deficits [3]. These drugs are also more effective if the patient is treated in the early stages of epilepsy; thus, a system for early detection of epilepsy can help to minimize the intake of antiepileptic drugs.

Various diagnostic methods have been proposed to detect epileptic seizures, the most common of which is electroencephalography (EEG). EEG signals acquired from healthy individuals are complex and chaotic. However, the complexity decreases during the onset of epileptic seizure. The ictogenesis mechanism that occurs in epilepsy involves transition from interictal state (before onset of seizure) to ictal state (after the onset of seizure), which leads to the decrease in complexity [3]. Various time and frequency analysis methods have been widely used to study epileptic disorders and their progression. Due to chaotic nature of EEG signals, nonlinear dynamics methods are preferred [4,5].

The correlation dimension (CD) method is widely used to study chaotic behaviors. In addition to biological signals from modalities such as EEG, electrocorticography (ECoG), and electromyography (EMG) [6,7], it can also be used to analyze seismic data for detecting the epicenter of earthquakes, financial market data, and other types of data [8–10]. Comparing the nonlinear dynamics of the interictal and ictal signals in epileptic disorders has attracted much attention [3,11–14]. The transition of a complex, irregular EEG signal in the interictal stage to a less complex signal in the ictal stage [3] is observed in CD as the ictal signals having a lower dimension than the interictal signals. Grassberger and Procaccia first proposed the fundamentals of the CD method [3]. In addition to using this method to study nonlinear dynamics in epileptic disorders, the original algorithm has been modified many times to study various signal characteristics [5,6].

In 1980, Takens proposed the theory embedding dimensions [15,16], which was later used by Grassberger and Procaccia to develop CD analysis [16,17]. They determined the CD by mapping a time-series signal into an $m$-dimensional embedding space. This $m$-dimensional embedding space with lag $L$ is given by

$$X_i = [x_i, x_{i+L}, \ldots, x_{i+(m-1)L}] \tag{1}$$

where $X_i$ is a reconstructed time-series vector of length $N$. Hence, $\{X_1, X_2, X_3, \ldots, X_N\}$ are vectors mapped into the $m$-dimensional embedding space. According to this conventional CD algorithm, the number of $(X_i, X_j)$ pairs separated by a distance $d_{ij}$ ($d_{ij} = |X_i - X_j|$) less than threshold $r$ is counted to obtain the correlation integral

$$C(N,r) = \frac{2}{(N)(N-1)} \sum_{i=1, i \neq j}^{N} H(r - d_{ij}) \tag{2}$$

Here, $H(r - d_{ij})$ represents the Heaviside function, which is expressed as

$$H(r - d_{ij}) = \begin{cases} 1; & d_{ij} \leq r \\ 0; & d_{ij} > r \end{cases} \tag{3}$$

Finally, CD is estimated from the slope of $\log C(N,r)$ versus $\log r$ with increasing embedding dimension $m$. Determining the saturation region and scaling region from this plot can be time consuming for a large amount of data. In addition, the scaling region may not fall within the plot, when the threshold required to determine the CD is large.

Normalizing the signal between 0 and 1 helps to generate the correlation integral using a smaller threshold [18–21]. However, it is important to choose an optimal threshold to determine the CD. A high threshold value makes CD sensitive to high amplitude spikes [13] and saturates the CD, because the majority of distances are considered in the estimation. When the threshold is too small, few distances contribute to the CD [22], making it unreliable. The CD can be used to analyze epileptic EEGs and to predict the onset of seizure. Maiwald *et al*. [12] designed a seizure prediction system that triggers an alarm when the CD crosses a specific threshold level, giving a warning several minutes before the onset of seizure. The drop in signal complexity occurs several minutes before the onset of seizure, which is known as the preictal state. During this state, Weinand *et al*. [23] observed increased blood flow to the epileptogenic lobe approximately 10 min before the onset of seizure. Aschenbrenner-Scheibe *et al.* [11] also observed a dimensionality drop during the preictal state, enabling them to predict seizure prior to onset.

Although the conventional CD algorithm has been widely used, our experiments revealed that it has some limitations. The correlation integral of signals is saturated at a small threshold, thus restricting our options for choosing the optimum threshold. Therefore, significant difference between interictal and ictal signals is achieved only for smaller threshold values. Moreover, using the Heaviside function to calculate the correlation integral yields only the total count of $(X_i, X_j)$ pairs with distances less than threshold $r$. Thus, the information about the actual distance between the vector pairs is lost. To retain this information, we modified the calculation of the correlation integral by taking the negative exponential of the distances less than the threshold. This modification enabled us to determine the correlation integral for the optimum range of threshold values and provided a larger difference in correlation integral between the interictal and ictal signals. The modified computation of the CD is explained in Section 2.

## 2. METHODS

In the modified correlation dimension (MCD) algorithm, the time-series signal embedded into *m*-dimensional embedding space with lag *L* is the same as equation 1. These vectors are mapped into the embedding space in the same way as in the conventional CD. Vectors $\{X_1, X_2, X_3, \dots, X_N\}$ in the embedding space are normalized by 1-norm normalization. The correlation integral is then modified as

$$C(N,r) = \frac{2}{(N)(N-1)} \sum_{i=1, i \neq j}^{N} P(r - d_{ij}) \qquad (4)$$

Here, $P(r - d_{ij})$ is defined as

$$P(r - d_{ij}) = \begin{cases} exp(-d_{ij}/r); & d_{ij} \leq r \\ 0; & d_{ij} > r \end{cases} \qquad (5)$$

where $0 < r < 1$ and $d_{ij} = |X_i - X_j|$.

In equation 4, the exponential of the distance between $(X_i, X_j)$ pairs less than threshold $r$ are summed to obtain the correlation integral. Normalizing the signal yields smaller $d_{ij}$ values; hence, they are amplified using a negative exponential function. Figure 1 shows a graphical comparison of $H(r - d_{ij})$ and $P(r - d_{ij})$. The value of the Heaviside function is 1 if the distance is less than $r$ and 0 otherwise. In contrast, $P(r - d_{ij})$ exponentially decays with increasing $d_{ij}$, and is set to zero above the threshold. Finally, the CD is obtained by determining the slope from the log-log plot of correlation integral versus threshold $r$.

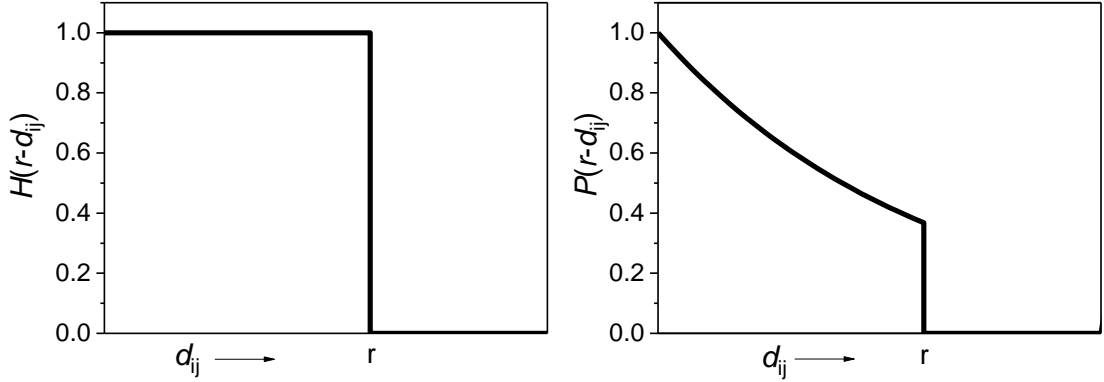

Fig 1. (Left) Heaviside function $H(r - d_{ij})$ and (right) modified function $P(r - d_{ij})$.

We used three datasets to test the MCD: a dataset comprising 100 pairs of interictal and ictal signals obtained from the Bonn University database [24]; a dataset comprising 50 pairs of focal epileptic and normal signals [14]; and a dataset comprising 14 pairs of interictal and ictal signals [25]. The signals from the datasets were normalized using the 1-norm method. The results for the last two datasets are shown in the Supplementary section. In the first dataset, the interictal and ictal signals were 23.6 s in duration with a sampling frequency of 173.61 Hz. We used a low-pass filter to obtain signals with a bandwidth of 0–60 Hz because frequency bands above 60 Hz mainly consist of noise. The correlation integral of the filtered signal is determined for $m = 1$ to 20 and lag of $L = 1$.

## 3. RESULTS AND DISCUSSION

Figure 2 shows the log-log plot of the correlation integral versus threshold for $m = 1$ to 20 using the CD and MCD methods. Compared with the CD method, the modified method gave a significantly larger difference between the ictal and interictal signals. An optimal threshold must be chosen to calculate the correlation integral. A small threshold fails to estimate significant distances between vectors, $d_{ij}$, whereas a large threshold will estimate all the distances between the vectors including noise. In this experiment, we compared correlation integrals for thresholds between $10^{-4}$ and 1, and obtained the best results for thresholds between $10^{-3}$ and $10^{-2}$. The correlation integral calculated using the CD method reaches saturation much faster than the one calculated using the modified method. Further information is provided in the Supplementary Information and Fig. S3, where we can see that slight alteration of the threshold values can result in a small difference between the ictal and interictal correlation integrals in the CD method. However, in the MCD method, all the selected threshold values yielded a good difference between the two (Fig. S3). Hence, in the MCD method, the difference in correlation integral between the interictal and ictal signals is much greater and is less sensitive to the choice of threshold. With the increase in embedding dimension $m$, we observed decreases in the slope for both the CD and MCD methods.

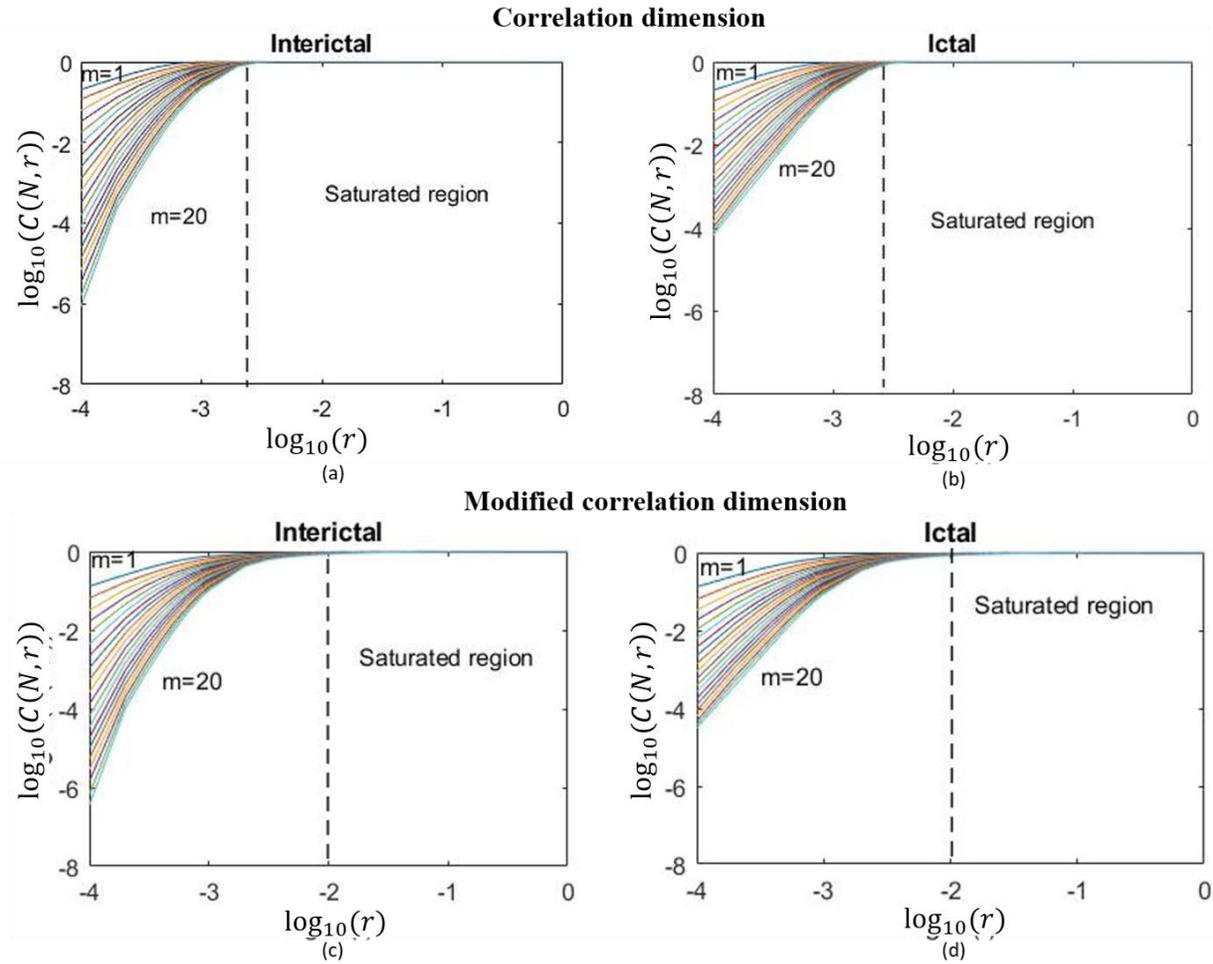

Fig 2. (a) Interictal and (b) ictal signals for the MCD method and (c) interictal and (d) ictal signals for the CD method.

Figure 3 shows heatmaps of $d_{ij}$ from the ictal and interictal signals using the CD and MCD methods. In both methods, the ictal signals consisted of a larger number of values closer to 0, indicating reduced complexity. Thus, we obtained a significantly lower correlation integral for ictal signals than for interictal signals. Furthermore, $H(r - d_{ij})$ used to estimate the correlation integral in the CD algorithm takes values of 0 or 1 and acts as a counter for estimating CD. In contrast, $P(r - d_{ij})$ in the MCD algorithm can take values of anywhere between 0 and 1, and thus uses the exact information from the signal to estimate the MCD.

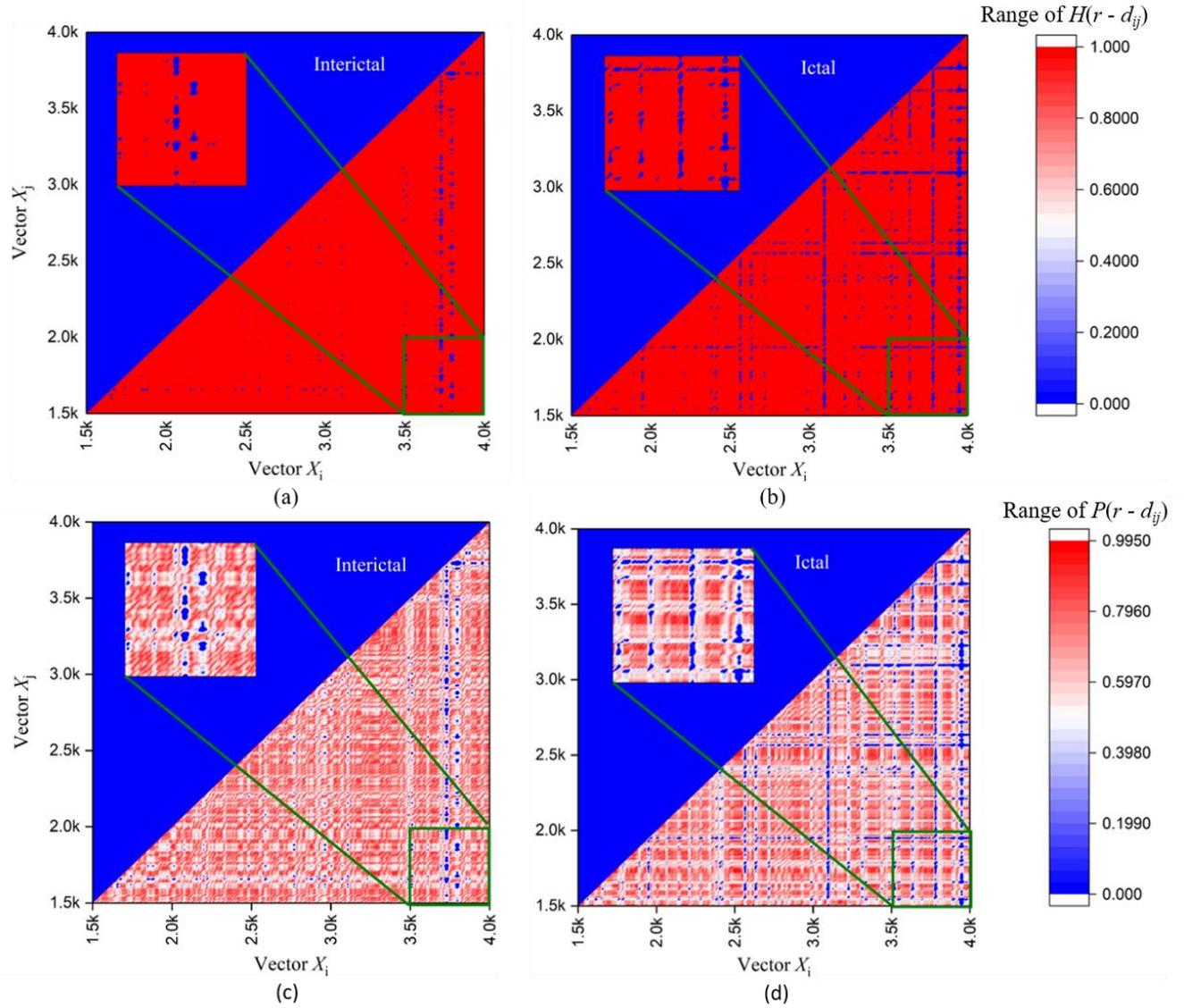

Fig 3. Heatmaps of $d_{ij}$ calculated using the CD method for (a) interictal and (b) ictal signals and using the MCD method for (c) interictal and (d) ictal signals.

For $m = 15$ and a threshold of 0.003, we computed the CD and MCD for 100 interictal and 100 ictal signals. CD and MCD both yielded a higher correlation integral for the interictal signals. Similar results were reported by Ying *et al*. [21]. They described a decrease in the correlation integral during the onset of seizure, which was due to the loss of the chaotic nature of the signal. The difference in the correlation integral between the interictal and ictal signals calculated by using the CD and MCD algorithms were compared. Figure 4 shows that the MCD method gave a higher average difference. Similar results were obtained with the other two datasets (Figs. S1 and S2). This result demonstrates that MCD is a more efficient algorithm than CD.

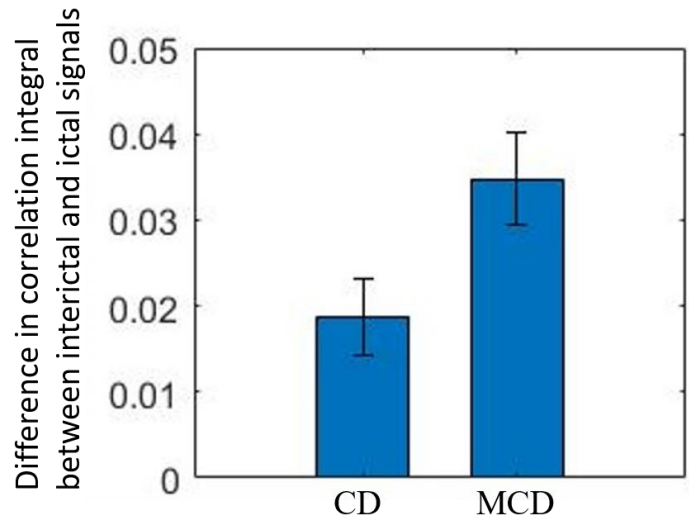

Fig. 4. Mean and standard error of difference in correlation integral between the interictal and ictal signals for the CD and MCD methods.

CONCLUSION

We modified the conventional CD algorithm for analyzing the interictal and ictal signals of epileptic patients to address its shortcomings. The CD algorithm uses the Heaviside function to determine the correlation integral. However, this algorithm generates a significant correlation integral value for only a small threshold. For a slightly higher threshold, the correlation integrals of both the interictal and ictal signals reach saturation, making them difficult to classify. Our MCD algorithm is not as sensitive as in the CD algorithm to the choice of threshold. For the datasets used in this experiment, we obtained a significant difference in the correlation integrals of the interictal and ictal signals at the optimal threshold value. This enabled us to classify the interictal and ictal signals more accurately using the MCD algorithm than with the CD algorithm. The correlation integrals showed that the MCD method outperformed the CD method. The superior performance of the MCD method in all three datasets demonstrated the generality of the new algorithm; hence, our method could also be used to study other neurological disorders.

# Modified Computation of Correlation Integral for Analyzing Epileptic Signals

Prajna Upadhyaya, Tohru Yagi

*Department of Mechanical Engineering, Tokyo Institute of Technology, 2-12-1 Meguro, Ookayama, Tokyo, 〒152-8550*

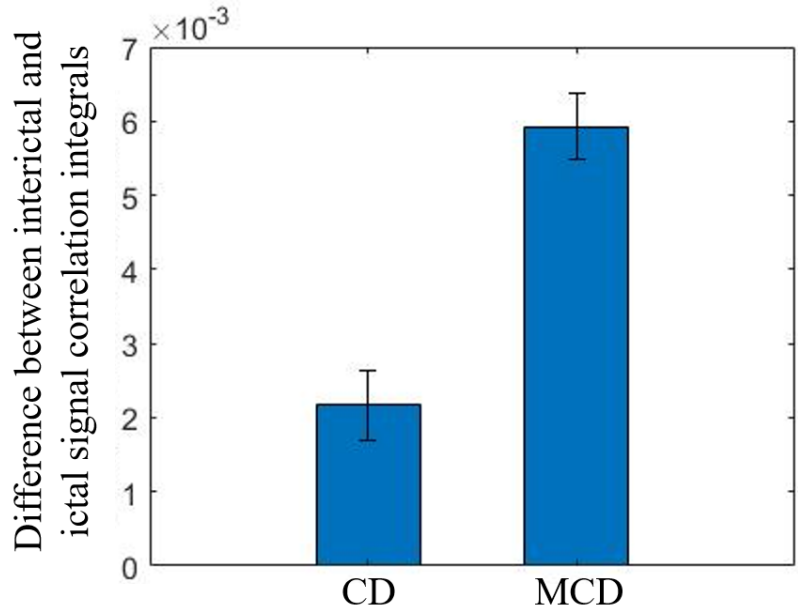

Fig. S1. Mean and standard error of difference between focal epileptic and normal signals.

To verify the generality of the improved efficiency of the MCD method compared with the CD method, we compared the results for a different set of 50 pairs of focal epileptic and normal signals taken from the Bonn University database (Ref. 14 in main paper). These signals have a sampling frequency of 512 Hz. The signals were low-pass filtered to obtain a bandwidth of 0–60 Hz. We compared the difference in correlation integral between focal epileptic and normal signals using both methods. Figure S1 shows that a larger mean difference was obtained between the focal epileptic and normal signals using the MCD method.

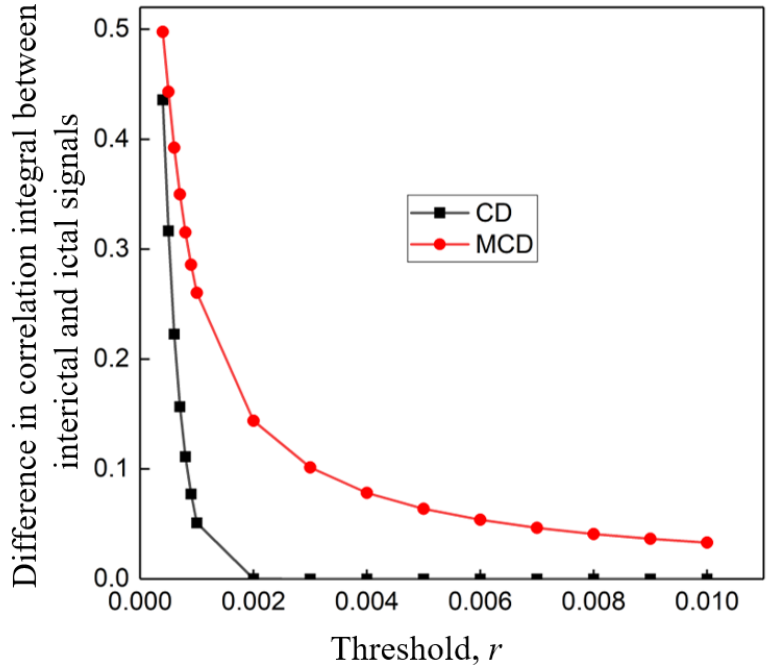

Fig. S2. Difference in correlation integral between interictal and ictal signals when varying the threshold value between 0.0005 and 0.01 using the (a) CD and (b) MCD methods.

The efficiency of the MCD method was also tested using the Temple University database (Ref. 25 in the main paper). We considered 40 s interictal and ictal epoch signals. We used 14 pairs of interictal and ictal signals with a sampling frequency of 250 Hz. A low-pass filter was used to obtain a bandwidth of 0–60 Hz. The CD and MCD algorithms were used to calculate the correlation integral, and the results were compared. Figure S2 shows that the CD algorithm can differentiate the interictal and ictal signals only for a small threshold, which limits our ability to choose the optimal threshold. However, the MCD algorithm allows a wider range of threshold to be used to differentiate the interictal and ictal signals. In addition, even at a smaller threshold, the MCD algorithm provides a larger difference between the interictal and ictal signals, and thus classifies them more accurately than the CD algorithm.

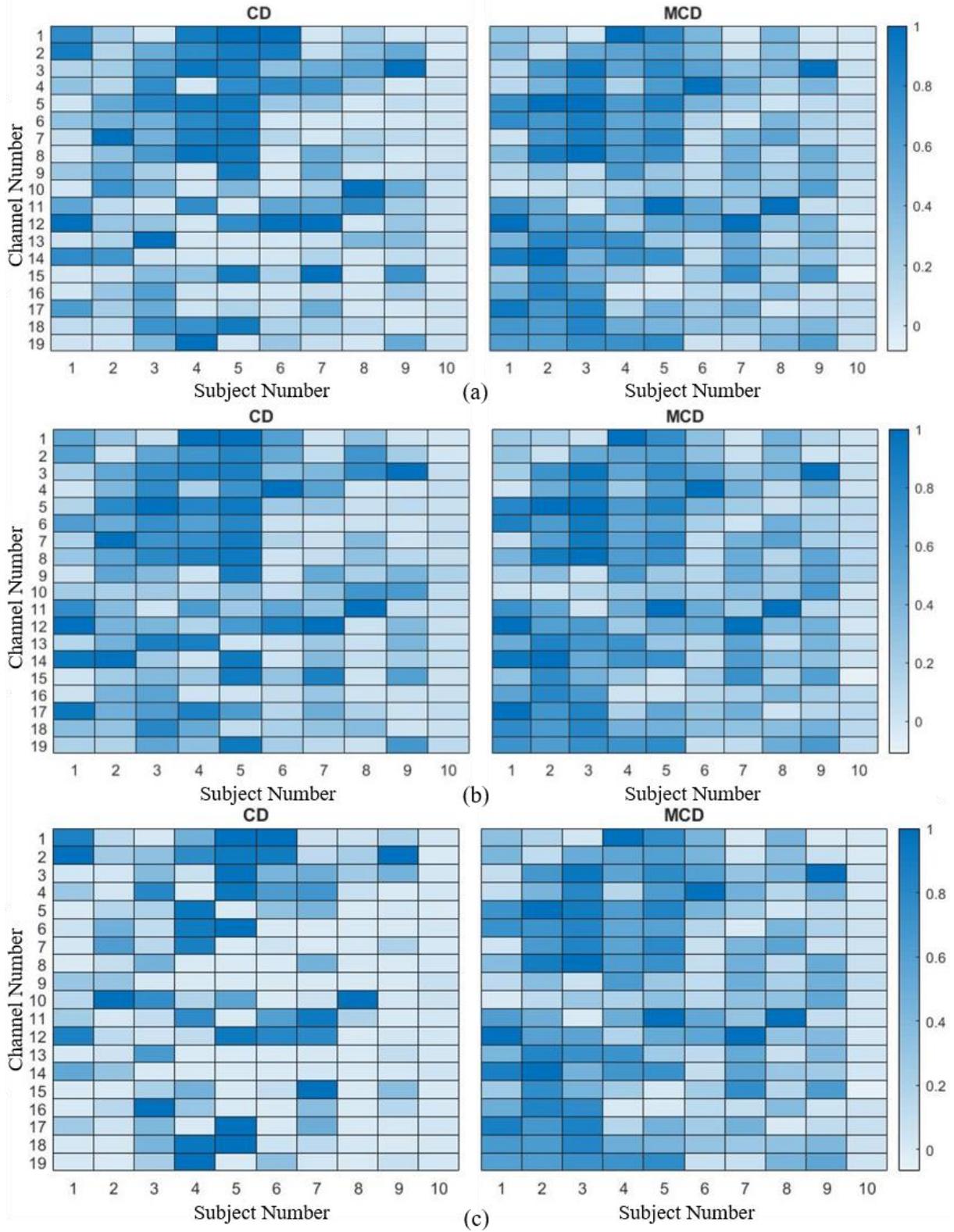

Fig. S3. Difference in correlation integral between the interictal and ictal signals for the CD and MCD methods for thresholds of (a) 0.0005, (b) 0.0007, and (c) 0.001.

Figure S3 shows heatmaps of the difference in correlation integral between the interictal and ictal signals obtained from the Temple University database for three thresholds. For thresholds of 0.001 and 0.0007, the performance of the CD algorithm was poor compared with the threshold 0.0005. In contrast, the performance of the MCD algorithm was good for all three thresholds, thereby confirming that the MCD algorithm outperforms the CD algorithm.